\documentclass[11pt,a4paper]{article}%
\usepackage{jheppub}
\usepackage{amsfonts}
\usepackage{amsmath}
\usepackage{amssymb}
\usepackage{graphicx}%
\pdfoutput=1

\begin{document}

\title{Einstein-AdS action, renormalized volume/area and holographic R\'{e}nyi entropies}
\author{Giorgos Anastasiou, Ignacio J. Araya, Cesar Arias and Rodrigo Olea}

\affiliation{Departamento de Ciencias F\'isicas, Universidad Andres Bello, \\ Sazi\'e 2212,
Piso 7, Santiago, Chile}

\emailAdd{georgios.anastasiou@unab.cl}
\emailAdd{araya.quezada.ignacio@gmail.com} \emailAdd{cesar.arias@unab.cl} \emailAdd{rodrigo.olea@unab.cl}

%\arxivnumber{1712.xxxxx}

\abstract{
We exhibit the equivalence between the renormalized volume of asymptotically
anti-de Sitter (AAdS) Einstein manifolds in four and six dimensions, and their
renormalized Euclidean bulk gravity actions. The action is that of Einstein
gravity, where the renormalization is achieved through the addition of a
single topological term. We generalize this equivalence, proposing an explicit
form for the renormalized volume of higher even-dimensional AAdS Einstein manifolds.
We also show that evaluating the renormalized bulk gravity action on the
conically singular manifold of the replica trick results in an action
principle that corresponds to the renormalized volume of the regular part of
the bulk, plus the renormalized area of a codimension-2 cosmic brane whose
tension is related to the replica index. Renormalized R\'{e}nyi entropy of
odd-dimensional holographic CFTs can thus be obtained from the renormalized
area of the brane with finite tension, including the effects of its
backreaction on the bulk geometry. The area computation corresponds to an
extremization problem for an enclosing surface that extends to the AdS
boundary, where the newly defined renormalized volume is considered.
}


\maketitle

\section{Introduction}

Asymptotically anti-de Sitter (AAdS) spaces are naturally endowed with a
conformal structure at their boundary
\cite{FG,Imbimbo:1999bj,Schwimmer:2008yh}, as evidenced from the divergent
conformal factor in the Fefferman-Graham (FG) expansion of the boundary metric
\cite{FG}. This divergence leads to an infinite volume, which in the context
of Einstein-AdS gravity corresponds to the infrared divergence of the
Einstein-Hilbert (EH) part of the action, being proportional to the volume of
the space when evaluated on-shell. In particular, for a $2n-$dimensional AAdS
manifold $M_{2n}$, we have that
\begin{equation}
I_{EH}\left[  M_{2n}\right]  =-\frac{\left(  2n-1\right)  }{8\pi G{\ell}^{2}%
}Vol\left[  M_{2n}\right]  , \label{RelationBare}%
\end{equation}
where ${\ell}$ is the AdS radius of the manifold and $Vol\left[
M_{2n}\right]  $ is its bare (divergent) volume. Therefore, it is reasonable
to expect that the problem of volume renormalization should be intimately
related to the renormalization of the Einstein-AdS action.

Volume renormalization has been thoroughly studied in the mathematical
literature. In the context of asymptotically hyperbolic (AH)
manifolds~\cite{2000math.....11051A,Yang2008,Alexakis2010,ALBIN2009140,Graham:1999pm,2016arXiv161209026F}%
, its resolution is known to provide relevant information about the boundary
geometry, in the form of conformal invariant quantities. More recently and by
means of tractor calculus techniques~\cite{Gover2003,bailey1994}, the
renormalized volume problem has been extended to more general setups
in~\cite{Gover:2016xwy,Gover:2016hqd}, where hidden algebraic structures are
enhanced by considering the bulk itself as a conformal manifold.

For the particular case of AAdS Einstein spaces, as discussed by Albin in
Ref.\cite{ALBIN2009140}, the renormalized volume can be expressed in terms of
the integral of some (unspecified) polynomial in contractions of the Weyl
curvature tensor, and the Euler characteristic of the manifold. Concrete
realizations of Albin's prescription exist in the cases of AAdS Einstein
manifolds in four \cite{2000math.....11051A} and six \cite{Yang2008}
dimensions. In the four-dimensional case, as discussed by Anderson in
Ref.\cite{2000math.....11051A}, the renormalized volume $Vol_{ren}\left[
M_{4}\right]  $ of the four-dimensional manifold $M_{4}$ obeys a relation
given by%

\begin{equation}
\frac{1}{32\pi^{2}}%
%TCIMACRO{\dint \limits_{M_{4}}}%
%BeginExpansion
{\displaystyle\int\limits_{M_{4}}}
%EndExpansion
d^{4}x\sqrt{G}W^{\alpha\beta\mu\nu}W_{\alpha\beta\mu\nu}=\chi\left[
M_{4}\right]  -\frac{3}{4\pi^{2}}Vol_{ren}\left[  M_{4}\right]  ,
\end{equation}
where $W$ is its Weyl tensor and $\chi\left[  M_{4}\right]  $ is its Euler characteristic.

The renormalization of the Einstein-AdS gravity action has been achieved
through the Holographic Renormalization procedure
\cite{Emparan:1999pm,Kraus:1999di,deHaro:2000vlm,Balasubramanian:1999re,Henningson:1998gx,Papadimitriou:2004ap,Papadimitriou:2005ii}%
, where one considers the addition of the Gibbons-Hawking-York term
\cite{York:1972sj,Gibbons:1976ue} and of intrinsic counterterms at the AdS
boundary. These boundary terms are fixed by requiring the cancellation of
divergences in the FG expansion of the action at the AdS boundary, while
maintaining consistency with a well-posed variational principle for the
conformal structure. However, as all the added counterterms are boundary
terms, it is difficult to see their relation to the renormalized volume, which
is defined in terms of a bulk integral. It is convenient, therefore, to
consider an alternative but equivalent renormalization scheme, named the
Kounterterms procedure
\cite{Olea:2005gb,Olea:2006vd,Kofinas:2007ns,Miskovic:2014zja,Miskovic:2009bm}%
, which is coordinate independent and considers the addition of topological
terms in order to achieve the renormalization of the action. In particular,
for the $2n-$dimensional case, the procedure considers the addition of the
$n-th$ Chern form, which corresponds to the boundary term arising from the
Euler theorem in $2n-$ dimensions. As shown in Ref.\cite{Miskovic:2014zja},
the renormalized Einstein-AdS action $I_{EH}^{ren}$, can be written precisely
in terms of the volume integral of a polynomial of antisymmetric contractions
of the Weyl tensor, hinting at its relation with the renormalized volume.

In this paper, we show that the relation between the EH part of the action and
the bare volume given by Eq.(\ref{RelationBare}), still holds between the
renormalized action $I_{EH}^{ren}\left[  M_{2n}\right]  $ and the renormalized
volume $Vol_{ren}\left[  M_{2n}\right]  $, for the cases where a definite
expression for the renormalized volume exists (the cases of AAdS manifolds in
four \cite{2000math.....11051A} and six \cite{Yang2008} dimensions). We also
conjecture that the relation holds for all even-dimensional AAdS Einstein
manifolds, and that therefore, the polynomial structure of the renormalized
action provides a concrete realization of Albin's prescription for the
renormalized volume.

In the context of the gauge/gravity duality
\cite{Maldacena:1997re,Gubser:1998bc,Witten:1998qj}, the volume
renormalization here discussed is also related to the renormalization
of\ holographic entanglement entropy (EE)
\cite{Ryu:2006bv,Rangamani:2016dms,Lewkowycz:2013nqa,Taylor:2016aoi} and
entanglement R\'{e}nyi entropy (ERE)
\cite{Dong:2016fnf,Headrick:2010zt,2011arXiv1102.2098B,Hung:2011nu}. This is
understood from the fact that both the EE and the ERE can be expressed in
terms of the areas of certain codimension-2 surfaces that are embedded within
an AAdS bulk manifold and which extend to the AdS boundary. Their
corresponding areas are infinite due to the same divergent conformal factor
that affects the bulk volume, and thus, they can be renormalized in a similar
way. Regarding the renormalization of EE, in Refs.
\cite{Anastasiou:2017xjr,Anastasiou:2018rla}, we developed a topological
scheme, which is applicable to holographic odd-dimensional conformal field
theories (CFTs) with even-dimensional AAdS gravity duals. The resulting
renormalized EE $\left(  S_{EE}^{ren}\right)  $ can be written as a polynomial
in contractions of the AdS curvature tensor $\mathcal{F}_{AdS}$ of a
codimension-2 surface $\Sigma$ in the AAdS bulk, and a purely topological term
that depends on its Euler characteristic $\chi\left[  \Sigma\right]  $. The
$\mathcal{F}_{AdS}$ tensor \cite{Mora:2006ka} is defined as%

\begin{equation}
\left.  \mathcal{F}_{AdS}\right.  _{b_{1}b_{2}}^{a_{1}a_{2}}=\mathcal{R}%
_{b_{1}b_{2}}^{a_{1}a_{2}}+\frac{1}{{\ell}^{2}}\delta_{\left[  b_{1}%
b_{2}\right]  }^{\left[  a_{1}a_{2}\right]  }, \label{FAdS}%
\end{equation}
where $\mathcal{R}_{b_{1}b_{2}}^{a_{1}a_{2}}$ is the intrinsic Riemann tensor
of $\Sigma$, ${\ell}$ is the bulk AdS radius and $\delta_{\left[  b_{1}%
b_{2}\right]  }^{\left[  a_{1}a_{2}\right]  }$ is the antisymmetric
generalization of the Kronecker delta. Also, considering the standard
Ryu-Takayanagi (RT) minimal area construction \cite{Ryu:2006bv}, the bulk
surface $\Sigma$ is the minimum of area (codimension-2 volume) which is
conformally cobordant with the entangling region in the CFT. As we show in
this paper, the interpretation of the renormalized EE as renormalized area is
then natural, as it has the same polynomial structure mentioned above, but
considering the AdS curvature of $\Sigma$ instead of the Weyl tensor of
$M_{2n}$. We therefore consider that%

\begin{equation}
S_{EE}^{ren}=\frac{Vol_{ren}\left[  \Sigma\right]  }{4G}\text{,}
\label{SEERen}%
\end{equation}
in analogy with the standard RT formula.

Beyond the renormalization of EE, it is of interest to study the
renormalization of ERE, as the latter is also expressible in terms of areas of
codimension-2 surfaces. As shown by Xi Dong, in \cite{Dong:2016fnf}, the
$m-th$ ERE $S_{m}$ can be written in terms of the integral of a quantity
called the modular entropy $\widetilde{S}_{m}$ \cite{Nishioka:2018khk}, which
in turn is obtained as the area of certain cosmic brane. In particular,
$\widetilde{S}_{m}$ is given by the area of a minimal cosmic brane $\Sigma
_{T}$ with tension $T\left(  m\right)  =\frac{\left(  m-1\right)  }{4mG}$,
such that it is conformally cobordant with the entangling region in the CFT.
The cosmic brane is minimal in the sense that it extremizes the total action
consisting on the Einstein-AdS action for the bulk manifold plus the
Nambu-Goto (NG) action for the brane, thus accounting for the backreaction of
the brane on the ambient geometry. $\widetilde{S}_{m}$ obeys an analog of the
RT area law, such that%

\begin{equation}
\widetilde{S}_{m}=\frac{Vol\left[  \Sigma_{T}\right]  }{4G}, \label{ModEnt}%
\end{equation}
and the $m-th$ R\'{e}nyi entropy $S_{m}$ can be computed as%

\begin{equation}
S_{m}=\frac{m}{m-1}%
%TCIMACRO{\dint \limits_{1}^{m}}%
%BeginExpansion
{\displaystyle\int\limits_{1}^{m}}
%EndExpansion
\frac{dm^{\prime}}{m^{\prime2}}\widetilde{S}_{m^{\prime}}, \label{RenyiEnt}%
\end{equation}
where the brane tension provides a natural analytic continuation of the
integer $m$ into the real numbers, which is required for performing the
integral. As we show in this paper, by considering the same volume
renormalization prescription discussed above, it is possible to obtain the
renormalized version of $\widetilde{S}_{m}$, given in terms of $Vol_{ren}%
\left[  \Sigma_{T}\right]  $, from which the renormalized ERE is computed in
the same way as in Eq.(\ref{RenyiEnt}). Also, a renormalized version of the
total action can derived by evaluating the renormalized Einstein-AdS action on
the conically singular bulk manifold that considers the brane as a singular
source in the Riemann curvature. Then, the contribution due to the cosmic
brane can be rewritten as a renormalized version of the NG action.

The organization of the paper is as follows: In Section \ref{Section II}, we
explicitate the relation between renormalized volume and renormalized
Einstein-AdS action, making contact with the mathematical literature in the
four \cite{2000math.....11051A} and six \cite{Yang2008} dimensional cases. We
also give our conjecture for the general relation in the $2n-$dimensional
case, relating it to Albin's prescription \cite{ALBIN2009140}. In Section
\ref{Section III}, we exhibit the emergence of the renormalized total action,
in agreement with Dong's cosmic brane prescription \cite{Dong:2016fnf}, from
evaluating the $I_{EH}^{ren}$ action on the conically singular manifold
considering the brane as a source. We also comment that the inclusion of the
renormalized NG action for the brane can be considered as a one-parameter
family of deformations in the definition of the renormalized bulk volume. In
Section \ref{Section IV}, we obtain the renormalized modular entropy
$\widetilde{S}_{m}^{ren}$ and the renormalized ERE $S_{m}^{ren}$ starting from
the renormalized total action, and we compare the resulting modular entropy
with the existing literature for renormalized areas of minimal surfaces
\cite{Alexakis2010}. In Section \ref{Section V}, we consider the computation
of $S_{m}^{ren}$ for the particular case of a ball-like entangling region at
the CFT$_{2n-1}$, and we check that the usual result for $S_{EE}^{ren}$ is
recovered in the $m\rightarrow1$ limit. Finally, in Section \ref{Section VI},
we comment on the physical applications of our conjectured renormalized volume
formula, on our topological procedure for computing renormalized EREs and on
future generalizations thereof.

\bigskip

\section{Renormalized Einstein-AdS action is renormalized
volume\label{Section II}}

\bigskip

The standard EH action, when evaluated on an AAdS Einstein manifold, is
proportional to the volume of the manifold, which is divergent. We propose
that for $2n-$dimensional spacetimes, the renormalized Einstein-AdS action
$I_{EH}^{ren}$ is also proportional to the corresponding renormalized volume
of the bulk manifold. To motivate this conjecture, we first introduce
$I_{EH}^{ren}$, and we then compare it with known formulas for the
renormalized volume of AAdS Einstein manifolds in four and six-dimensions.
Finally, we give a concrete formula for the renormalized volume in the general
$2n-$dimensional case, and we comment on its properties.

There are different but equivalent prescriptions for renormalizing the action.
For example, the standard Holographic Renormalization scheme
\cite{Emparan:1999pm,Kraus:1999di,deHaro:2000vlm,Balasubramanian:1999re,Henningson:1998gx,Papadimitriou:2004ap,Papadimitriou:2005ii}
and the Kounterterms procedure
\cite{Olea:2005gb,Olea:2006vd,Kofinas:2007ns,Miskovic:2014zja,Miskovic:2009bm}%
. The equivalence between the two renormalization schemes, for the case of
Einstein-AdS gravity, was explicitly discussed in
Refs.\cite{Miskovic:2014zja,Miskovic:2009bm}, so using either one or the other
is a matter of convenience. However, as discussed in the introduction, we
consider the Kounterterms-renormalized action as it can be readily compared
with the existing renormalized volume formulas.

We consider the $2n-$ dimensional Einstein-AdS action $I_{EH}^{ren}$ as
derived using the Kounterterms prescription \cite{Olea:2005gb}, which is given
by
\begin{equation}
I_{EH}^{ren}\left[  M_{2n}\right]  =\frac{1}{16\pi G}\displaystyle\int
\limits_{M_{2n}}d^{2n}x\sqrt{G}\left(  R-2\Lambda\right)  +\frac{c_{2n}}{16\pi
G}\displaystyle\int\limits_{\partial M_{2n}}B_{2n-1}, \label{IRenChern}%
\end{equation}
where $c_{2n}$ is defined as%

\begin{equation}
c_{2n}=\frac{\left(  -1\right)  ^{n}{\ell}^{2n-2}}{n\left(  2n-2\right)  !}
\label{c_2n}%
\end{equation}
and $B_{2n-1}$ is the $n-th$ Chern form, which in Gauss normal coordinates
(considering a radial foliation) corresponds to%

\begin{align}
B_{2n-1}\overset{\text{def}}{=}-2n\displaystyle\int\limits_{0}^{1}dtd^{2n-1}x
&  \sqrt{h}\delta_{\left[  i_{1}\cdots i_{2n-1}\right]  }^{\left[  j_{1}\cdots
j_{2n-1}\right]  }K_{j_{1}}^{i_{1}}\left(  \frac{1}{2}\mathcal{R}_{j_{2}j_{3}%
}^{i_{2}i_{3}}-t^{2}K_{j_{2}}^{i_{2}}K_{j_{3}}^{i_{3}}\right)  \times
\nonumber\\
&  \cdots\times\left(  \frac{1}{2}\mathcal{R}_{j_{2n-2}j_{2n-1}}%
^{i_{2n-2}i_{2n-1}}-t^{2}K_{j_{2n-2}}^{i_{2n-2}}K_{j_{2n-1}}^{i_{2n-1}%
}\right)  . \label{Chern}%
\end{align}
In Eq.(\ref{Chern}), $\mathcal{R}_{j_{1}j_{2}}^{i_{1}i_{2}}$ is the intrinsic
Riemann tensor at the AdS boundary and $K_{j}^{i}$ is its extrinsic curvature
with respect to the radial foliation. We emphasize that the addition of the
$B_{2n-1}$ term, which has an explicit dependence on the extrinsic curvature
$K$, is consistent with a well-posed variational principle with Dirichlet
boundary conditions for the CFT metric $g_{\left(  0\right)  }$, instead of
the usual Dirichlet condition for the induced metric $h$. In
particular, considering the FG expansion, variations of $K$ are proportional to
variations of $g_{\left(  0\right)  }$.

As shown in Ref.\cite{Miskovic:2014zja}, the renormalized action $I_{EH}%
^{ren}\left[  M_{2n}\right]  $ can also be written as%

\begin{equation}
I_{EH}^{ren}\left[  M_{2n}\right]  =\frac{1}{16\pi G}\displaystyle\int
\limits_{M_{2n}}d^{2n}x\sqrt{G}\left(  {\ell}^{2\left(  n-1\right)  }%
P_{2n}\left[  W_{\left(  E\right)  }\right]  \right)  -\frac{c_{2n}}{16\pi
G}\left(  4\pi\right)  ^{n}n!\chi\left[  M_{2n}\right]  , \label{IRenPoly}%
\end{equation}
where $P_{2n}\left[  X\right]  $ is a polynomial of a rank $\binom{2}{2}%
-$tensor $X$, given by%

\begin{equation}
P_{2n}\left[  X\right]  =\frac{1}{2^{n}n\left(  2n-2\right)  !}%
%TCIMACRO{\dsum \limits_{k=1}^{n}}%
%BeginExpansion
{\displaystyle\sum\limits_{k=1}^{n}}
%EndExpansion
\frac{\left(  -1\right)  ^{k}\left[  2\left(  n-k\right)  \right]  !2^{\left(
n-k\right)  }}{{\ell}^{2\left(  n-k\right)  }}\binom{n}{k}\delta_{\left[
\mu_{1}...\mu_{2k}\right]  }^{\left[  \nu_{1}...\nu_{2k}\right]  }X_{\nu
_{1}\nu_{2}}^{\mu_{1}\mu_{2}}\cdots X_{\nu_{2k-1}\nu_{2k}}^{\mu_{2k-1}\mu
_{2k}}, \label{P2n}%
\end{equation}
$\delta_{\left[  \nu_{1}...\nu_{k}\right]  }^{\left[  \mu_{1}...\mu
_{k}\right]  }$ is the totally antisymmetric generalization of the Kronecker
delta defined as $\delta_{\left[  \nu_{1}...\nu_{k}\right]  }^{\left[  \mu
_{1}...\mu_{k}\right]  }=\det\left[  \delta_{\nu_{1}}^{\mu_{1}}\cdots
\delta_{\nu_{k}}^{\mu_{k}}\right]  $ and $W_{\left(  E\right)  }$ is the Weyl
tensor of an AAdS Einstein manifold, which can be written as%

\begin{equation}
\left.  W_{\left(  E\right)  }\right.  _{\nu_{1}\nu_{2}}^{\mu_{1}\mu_{2}%
}=R_{\nu_{1}\nu_{2}}^{\mu_{1}\mu_{2}}+\frac{1}{{\ell}^{2}}\delta_{\left[
\nu_{1}\nu_{2}\right]  }^{\left[  \mu_{1}\mu_{2}\right]  }. \label{Weyl(E)}%
\end{equation}
In order to show the equivalence between Eq.(\ref{IRenChern}) and
Eq.(\ref{IRenPoly}), one starts by considering the $2n-$dimensional Euler
theorem, which relates the $n-th$ boundary Chern form with the $2n-$%
dimensional Euler density $\mathcal{E}_{2n}$. In particular, one has that%

\begin{equation}
\displaystyle\int\limits_{M_{2n}}\mathcal{E}_{2n}=\left(  4\pi\right)
^{n}n!\chi\left[  M_{2n}\right]  +\displaystyle\int\limits_{\partial M_{2n}%
}B_{2n-1}, \label{EulerTheo}%
\end{equation}
where $\mathcal{E}_{2n}$ is given in terms of the bulk Riemann tensor by%

\begin{equation}
\mathcal{E}_{2n}=\frac{1}{2^{n}}d^{2n}x\sqrt{G}\delta_{\left[  \mu_{1}%
...\mu_{2n}\right]  }^{\left[  \nu_{1}...\nu_{2n}\right]  }R_{\nu_{1}\nu_{2}%
}^{\mu_{1}\mu_{2}}\cdots R_{\nu_{2n-1}\nu_{2n}}^{\mu_{2n-1}\mu_{2n}},
\label{EulerDensity}%
\end{equation}
and $\chi\left[  M_{2n}\right]  $ is the Euler characteristic of the bulk
manifold. One can then re-write the $n-th$ Chern form appearing in
Eq.(\ref{IRenChern}) in terms of $\mathcal{E}_{2n}$ and considering that for
AAdS Einstein manifolds%

\begin{equation}
R_{\nu_{1}\nu_{2}}^{\mu_{1}\mu_{2}}=\left.  W_{\left(  E\right)  }\right.
_{\nu_{1}\nu_{2}}^{\mu_{1}\mu_{2}}-\frac{1}{{\ell}^{2}}\delta_{\left[  \nu
_{1}\nu_{2}\right]  }^{\left[  \mu_{1}\mu_{2}\right]  },
\end{equation}
it is possible to show that%

\begin{equation}
\displaystyle\int\limits_{M_{2n}}d^{2n}x\sqrt{G}{\ell}^{2\left(  n-1\right)
}P_{2n}\left[  W_{\left(  E\right)  }\right]  =\displaystyle\int
\limits_{M_{2n}}d^{2n}x\sqrt{G}\left(  R-2\Lambda\right)  +c_{2n}%
\displaystyle\int\limits_{M_{2n}}\mathcal{E}_{2n}, \label{PtoE}%
\end{equation}
with%

\begin{equation}
\Lambda=-\frac{\left(  2n-1\right)  \left(  2n-2\right)  }{2{\ell}^{2}%
},~R=-\frac{2n\left(  2n-1\right)  }{{\ell}^{2}},\ \ \ \ \ \label{LandR}%
\end{equation}
$c_{2n}$ as given in Eq.(\ref{c_2n}) and $P_{2n}\left[  X\right]  $ as given
in Eq.(\ref{P2n}), making use of the properties of the generalized Kronecker
delta functions. Then, the equivalence between Eq.(\ref{IRenChern}) and
Eq.(\ref{IRenPoly}) follows trivially. We note that $I_{EH}^{ren}\left[
M_{2n}\right]  $ as given in Eq.(\ref{IRenPoly}), when considering that
$tr[W_{\left(  E\right)  }]=0$ for Einstein manifolds, only differs from the
renormalized action given in Ref.\cite{Miskovic:2014zja} by the constant term
proportional to $\chi\left[  M_{2n}\right]  $. However, as discussed in
Ref.\cite{Olea:2005gb}, this term does not alter the dynamics of the equations
of motion, nor changes the overall thermodynamic properties of the solutions,
and it only amounts to a trivial shift in the value of the horizon entropy for
black-hole solutions.

Considering the form of $I_{EH}^{ren}\left[  M_{2n}\right]  $ given in
Eq.(\ref{IRenPoly}), we now show that, as mentioned in the introduction, the
renormalized volume of AAdS Einstein manifolds in four and six dimensions is
indeed proportional to the renormalized Einstein-AdS action, with the same
proportionality factor considered in Eq.(\ref{RelationBare}). These cases
serve as examples for our conjectured relation in the general $2n-$dimensional
case, which we discuss afterwards.

\subsection{The 4D case}

In the case of four-dimensional AAdS manifolds, as seen from setting $n=2$ in
Eq.(\ref{IRenPoly}), the renormalized Einstein-AdS action is given by%

\begin{equation}
I_{EH}^{ren}\left[  M_{4}\right]  =\frac{{\ell}^{2}}{256\pi G}%
%TCIMACRO{\dint \limits_{M_{4}}}%
%BeginExpansion
{\displaystyle\int\limits_{M_{4}}}
%EndExpansion
d^{4}x\sqrt{G}\delta_{\left[  \mu_{1}\mu_{2}\mu_{3}\mu_{4}\right]  }^{\left[
\nu_{1}\nu_{2}\nu_{3}\nu_{4}\right]  }\left.  W_{\left(  E\right)  }\right.
_{\nu_{1}\nu_{2}}^{\mu_{1}\mu_{2}}\left.  W_{\left(  E\right)  }\right.
_{\nu_{3}\nu_{4}}^{\mu_{3}\mu_{4}}-\frac{\pi{\ell}^{2}}{2G}\chi\left[
M_{4}\right]  . \label{I4D}%
\end{equation}
Now, considering the definition of $\left\vert W_{\left(  E\right)
}\right\vert ^{2}$ as%

\begin{equation}
\left\vert W_{\left(  E\right)  }\right\vert ^{2}\overset{\text{def}}%
{=}\left.  W_{\left(  E\right)  }\right.  ^{\alpha\beta\mu\nu}\left.
W_{\left(  E\right)  }\right.  _{\alpha\beta\mu\nu},
\end{equation}
and that%

\begin{equation}
\frac{1}{16}%
%TCIMACRO{\dint \limits_{M_{4}}}%
%BeginExpansion
{\displaystyle\int\limits_{M_{4}}}
%EndExpansion
d^{4}x\sqrt{G}\delta_{\left[  \mu_{1}\mu_{2}\mu_{3}\mu_{4}\right]  }^{\left[
\nu_{1}\nu_{2}\nu_{3}\nu_{4}\right]  }\left.  W_{\left(  E\right)  }\right.
_{\nu_{1}\nu_{2}}^{\mu_{1}\mu_{2}}\left.  W_{\left(  E\right)  }\right.
_{\nu_{3}\nu_{4}}^{\mu_{3}\mu_{4}}=\frac{1}{4}%
%TCIMACRO{\dint \limits_{M_{4}}}%
%BeginExpansion
{\displaystyle\int\limits_{M_{4}}}
%EndExpansion
d^{4}x\sqrt{G}\left\vert W_{\left(  E\right)  }\right\vert ^{2},
\end{equation}
$I_{EH}^{ren}\left[  M_{4}\right]  $ can be re-written as%

\begin{equation}
I_{EH}^{ren}\left[  M_{4}\right]  =\frac{{\ell}^{2}}{64\pi G}%
%TCIMACRO{\dint \limits_{M_{4}}}%
%BeginExpansion
{\displaystyle\int\limits_{M_{4}}}
%EndExpansion
d^{4}x\sqrt{G}\left\vert W_{\left(  E\right)  }\right\vert ^{2}-\frac{\pi
{\ell}^{2}}{2G}\chi\left[  M_{4}\right]  .
\end{equation}
Finally, our proposal for the renormalized volume is given by%

\begin{equation}
Vol_{ren}\left[  M_{4}\right]  =-\frac{8\pi G{\ell}^{2}}{3}I_{EH}^{ren}\left[
M_{4}\right]  , \label{RenVol4}%
\end{equation}
and so we have that%

\begin{equation}
\frac{1}{32\pi^{2}}%
%TCIMACRO{\dint \limits_{M_{4}}}%
%BeginExpansion
{\displaystyle\int\limits_{M_{4}}}
%EndExpansion
d^{4}x\sqrt{G}\left\vert W_{\left(  E\right)  }\right\vert ^{2}=\chi\left[
M_{4}\right]  -\frac{3}{4\pi^{2}{\ell}^{4}}Vol_{ren}\left[  M_{4}\right]  ,
\end{equation}
in agreement with Anderson's formula \cite{2000math.....11051A}. By
considering Eq.(\ref{RenVol4}), we then conclude that in four dimensions the
renormalized Einstein-AdS action is indeed proportional to the renormalized
volume, and it is trivial to check that the proportionality factor is the same
as the one between the EH part of the action and the unrenormalized volume.

\subsection{The 6D case\label{2.2}}

In the case of six-dimensional AAdS manifolds, as seen from setting $n=3$ in
Eq.(\ref{IRenPoly}), the renormalized Einstein-AdS action is given by%

\begin{equation}
I_{EH}^{ren}\left[  M_{6}\right]  =\frac{1}{16\pi G}%
%TCIMACRO{\dint \limits_{M_{6}}}%
%BeginExpansion
{\displaystyle\int\limits_{M_{6}}}
%EndExpansion
d^{6}x\sqrt{G}{\ell}^{4}P_{6}\left[  W_{\left(  E\right)  }\right]  +\frac
{\pi^{2}{\ell}^{4}}{3G}\chi\left[  M_{6}\right]  , \label{I6D}%
\end{equation}
where the $P_{6}$ polynomial in contractions of the Weyl tensor is given by%

\begin{align}
P_{6}\left[  W_{\left(  E\right)  }\right]   &  =\frac{1}{2\left(  4!\right)
{\ell}^{2}}\delta_{\left[  \mu_{1}\mu_{2}\mu_{3}\mu_{4}\right]  }^{\left[
\nu_{1}\nu_{2}\nu_{3}\nu_{4}\right]  }\left.  W_{\left(  E\right)  }\right.
_{\nu_{1}\nu_{2}}^{\mu_{1}\mu_{2}}\left.  W_{\left(  E\right)  }\right.
_{\nu_{3}\nu_{4}}^{\mu_{3}\mu_{4}}\nonumber\\
&  -\frac{1}{\left(  4!\right)  ^{2}}\delta_{\left[  \mu_{1}\mu_{2}\mu_{3}%
\mu_{4}\mu_{5}\mu_{6}\right]  }^{\left[  \nu_{1}\nu_{2}\nu_{3}\nu_{4}\nu
_{5}\nu_{6}\right]  }\left.  W_{\left(  E\right)  }\right.  _{\nu_{1}\nu_{2}%
}^{\mu_{1}\mu_{2}}\left.  W_{\left(  E\right)  }\right.  _{\nu_{3}\nu_{4}%
}^{\mu_{3}\mu_{4}}\left.  W_{\left(  E\right)  }\right.  _{\nu_{5}\nu_{6}%
}^{\mu_{5}\mu_{6}}. \label{P6}%
\end{align}
Therefore, the Euler characteristic of the bulk manifold $M_{2n}$ can be
written as%

\begin{equation}
\chi\left[  M_{6}\right]  =\frac{3G}{\pi^{2}{\ell}^{4}}I_{EH}^{ren}\left[
M_{6}\right]  -\frac{3}{16\pi^{3}}%
%TCIMACRO{\dint \limits_{M_{6}}}%
%BeginExpansion
{\displaystyle\int\limits_{M_{6}}}
%EndExpansion
d^{6}x\sqrt{G}P_{6}\left[  W_{\left(  E\right)  }\right]  .
\end{equation}
\qquad

Now, we can rewrite $P_{6}\left[  W_{\left(  E\right)  }\right]  $ in terms of
$\left\vert W_{\left(  E\right)  }\right\vert ^{2}$ and the Weyl invariants in
six dimensions (setting ${\ell=1}$ for simplicity). In order to do this, we
consider that the first two Weyl invariants, $I_{1}$ and $I_{2}$, are given by%

\begin{align}
I_{1}\left[  W\right]   &  =W_{\alpha\beta\mu\nu}W^{\alpha\rho\lambda\nu
}W_{\rho\text{\ \ ~~~}\lambda}^{\text{ \ }\beta\mu},\nonumber\\
I_{2}\left[  W\right]   &  =W_{\mu\nu}^{\alpha\beta}W_{\alpha\beta}%
^{\rho\lambda}W_{\rho\lambda}^{\mu\nu},\ \ \ \ \label{WeylInvs}%
\end{align}
where $W$ denotes the Weyl tensor of a manifold that need not be Einstein.
Then, considering that%
\begin{align}
\delta_{\left[  \nu_{1}\nu_{2}\nu_{3}\nu_{4}\right]  }^{\left[  \mu_{1}\mu
_{2}\mu_{3}\mu_{4}\right]  }\left.  W_{\left(  E\right)  }\right.  _{\mu
_{1}\mu_{2}}^{\nu_{1}\nu_{2}}\left.  W_{\left(  E\right)  }\right.  _{\mu
_{3}\mu_{4}}^{\nu_{3}\nu_{4}}  &  =4\left\vert W_{\left(  E\right)
}\right\vert ^{2},\nonumber\\
\delta_{\left[  \nu_{1}\nu_{2}\nu_{3}\nu_{4}\nu_{5}\nu_{6}\right]  }^{\left[
\mu_{1}\mu_{2}\mu_{3}\mu_{4}\mu_{5}\mu_{6}\right]  }\left.  W_{\left(
E\right)  }\right.  _{\mu_{1}\mu_{2}}^{\nu_{1}\nu_{2}}\left.  W_{\left(
E\right)  }\right.  _{\mu_{3}\mu_{4}}^{\nu_{3}\nu_{4}}\left.  W_{\left(
E\right)  }\right.  _{\mu_{5}\mu_{6}}^{\nu_{5}\nu_{6}}  &  =16\left(
2I_{2}\left[  W_{\left(  E\right)  }\right]  +4I_{1}\left[  W_{\left(
E\right)  }\right]  \right)  ,
\end{align}
we obtain
\begin{equation}
-4!P_{6}\left[  W_{\left(  E\right)  }\right]  =-2\left\vert W_{\left(
E\right)  }\right\vert ^{2}+\frac{7}{3}I_{2}\left[  W_{\left(  E\right)
}\right]  -\frac{4}{3}I_{1}\left[  W_{\left(  E\right)  }\right]  +\left(
4I_{1}\left[  W_{\left(  E\right)  }\right]  -I_{2}\left[  W_{\left(
E\right)  }\right]  \right)  .
\end{equation}
Now, we consider an identity given by Osborn and Stergiou in
Ref.\cite{Osborn:2015rna}, which for Einstein manifolds in six dimensions
states that%

\begin{equation}
4I_{1}\left[  W_{\left(  E\right)  }\right]  -I_{2}\left[  W_{\left(
E\right)  }\right]  =\left.  W_{\left(  E\right)  }\right.  ^{\rho\mu
\nu\lambda}\square\left.  W_{\left(  E\right)  }\right.  _{\rho\mu\nu\lambda
}+10\left\vert W_{\left(  E\right)  }\right\vert ^{2}, \label{IdentityOsborn}%
\end{equation}
where $\square\overset{\text{def}}{=}\nabla_{\mu}\nabla^{\mu}$ is the
covariant Laplacian operator (for more details on the identity, see Appendix
\ref{Appendix A}). Then, integrating by parts, we have
\begin{equation}
4I_{1}\left[  W_{\left(  E\right)  }\right]  -I_{2}\left[  W_{\left(
E\right)  }\right]  =-\left\vert \nabla W_{\left(  E\right)  }\right\vert
^{2}+10\left\vert W_{\left(  E\right)  }\right\vert ^{2}+\text{b.t.},
\label{BTEq}%
\end{equation}
where b.t. is a boundary term that in our case plays no role, as it vanishes
asymptotically near the AdS boundary, and is therefore neglected (see Appendix
\ref{Appendix B}). Then, we obtain
\begin{equation}
-4!P_{6}=-\left\vert \nabla W_{\left(  E\right)  }\right\vert ^{2}+8\left\vert
W_{\left(  E\right)  }\right\vert ^{2}+\frac{7}{3}I_{2}\left[  W_{\left(
E\right)  }\right]  -\frac{4}{3}I_{1}\left[  W_{\left(  E\right)  }\right]  .
\end{equation}
Finally, we consider the definition of the conformal invariant $J$, given by
Chang, Quing and Yang in Ref.\cite{Yang2008} as%

\begin{equation}
J\left[  W\right]  =-\left\vert \nabla W\right\vert ^{2}+8\left\vert
W\right\vert ^{2}+\frac{7}{3}W_{\mu\nu}^{\text{ \ \ }\alpha\beta}%
W_{\alpha\beta}^{\text{ \ \ \ }\lambda\rho}W_{\lambda\rho}^{\text{ \ \ \ }%
\mu\nu}+\frac{4}{3}W_{\mu\nu\rho\lambda}W^{\mu\alpha\rho\beta}W_{\text{
~}\alpha\text{ \ ~~}\beta}^{\nu~~\lambda},
\end{equation}
and we see that%

\begin{equation}
P_{6}\left[  W_{\left(  E\right)  }\right]  =-\frac{1}{4!}J\left[  W_{\left(
E\right)  }\right]  .
\end{equation}
Therefore, the Euler characteristic of the $M_{6}$ bulk manifold can be
written as%

\begin{equation}
\chi\left[  M_{6}\right]  =\frac{3G}{\pi^{2}{\ell}^{4}}I_{EH}^{ren}\left[
M_{6}\right]  -\frac{1}{128\pi^{3}}%
%TCIMACRO{\dint \limits_{M_{6}}}%
%BeginExpansion
{\displaystyle\int\limits_{M_{6}}}
%EndExpansion
d^{6}x\sqrt{G}J\left[  W_{\left(  E\right)  }\right]  ,
\end{equation}
and by considering that, according to our proposal for renormalized volume,%

\begin{equation}
I_{EH}^{ren}\left[  M_{6}\right]  =-\frac{5}{8\pi G{\ell}^{2}}Vol_{ren}\left[
M_{6}\right]  , \label{RenVol6}%
\end{equation}
we have that%

\begin{equation}
\chi\left[  M_{6}\right]  =-\frac{15}{8\pi^{3}{\ell}^{6}}Vol_{ren}\left[
M_{6}\right]  +\frac{1}{128\pi^{3}}%
%TCIMACRO{\dint \limits_{M_{6}}}%
%BeginExpansion
{\displaystyle\int\limits_{M_{6}}}
%EndExpansion
d^{6}x\sqrt{G}J\left[  W_{\left(  E\right)  }\right]  ,
\end{equation}
in agreement with the renormalized volume formula proposed by Chang, Qing and
Yang \cite{Yang2008}. Inspection of Eq.(\ref{RenVol6}) allows us to conclude
that, in the six-dimensional case, the renormalized action is indeed
proportional to the renormalized volume, where the proportionality factor is
again the same as between the EH part of the action and the unrenormalized volume.

\subsection{The general even-dimensional case}

In agreement with the four-dimensional and six-dimensional cases, we propose
that the renormalized volume of $2n-$dimensional AAdS Einstein manifolds is
proportional to the renormalized Einstein-AdS action, such that%

\begin{equation}
I_{EH}^{ren}\left[  M_{2n}\right]  =-\frac{\left(  2n-1\right)  }{8\pi G{\ell
}^{2}}Vol_{ren}\left[  M_{2n}\right]  . \label{I_ren/Vol_ren}%
\end{equation}
Considering this proposal and the polynomial form of the renormalized action
as presented in Eq.(\ref{IRenPoly}), we conjecture that the renormalized
volume of $M_{2n}$ is given by%

\begin{equation}
Vol_{ren}\left[  M_{2n}\right]  =-\frac{{\ell}^{2}}{2\left(  2n-1\right)
}\left(
%TCIMACRO{\dint \limits_{M_{2n}}}%
%BeginExpansion
{\displaystyle\int\limits_{M_{2n}}}
%EndExpansion
d^{2n}x\sqrt{G}{\ell}^{2\left(  n-1\right)  }P_{2n}\left[  W_{\left(
E\right)  }\right]  -c_{2n}\left(  4\pi\right)  ^{n}n!\chi\left[
M_{2n}\right]  \right)  , \label{VolRenM2n}%
\end{equation}
where $c_{2n}$ and $W_{\left(  E\right)  }$ are defined in Eq.(\ref{c_2n}) and
Eq.(\ref{Weyl(E)}) respectively.

We note that our expression for $Vol_{ren}\left[  M_{2n}\right]  $ corresponds
to a concrete realization of Albin's prescription, given in
Ref.\cite{ALBIN2009140}, which considers that for even-dimensional AAdS
Einstein spaces, the renormalized volume can be expressed in terms of the
integral of a polynomial in contractions of the Weyl curvature tensor, and the
Euler characteristic of the manifold. Also, the obtained expression for the
renormalized volume has the same form as the expression for the renormalized
EE of holographic CFTs obtained in Ref.\cite{Anastasiou:2018rla}, multiplied
by $4G$. Therefore, both instances of renormalized volumes (for both the bulk
manifold $M_{2n}$ and the minimal surface $\Sigma$ of the RT construction) can
be put in the same footing. We also note that the renormalized volume
expression is trivial for constant curvature AdS manifolds. In particular, in
the constant curvature case, the Weyl tensor $W_{\left(  E\right)  }$ of
$M_{2n}$ vanishes identically, and so the only remaining contribution to the
renormalized volume is the purely topological constant, which is proportional
to the Euler characteristic of the manifold. The renormalized volume can
therefore be understood as a measure of the deviation of a manifold with
respect to the constant curvature case, which corresponds to the maximally
symmetric vacuum of the Einstein-AdS gravity theory (usually global AdS).

We will see in the following sections that the renormalized volume formula is
also applicable to codimension-2 surfaces that minimize a certain total action
which corresponds to the renormalized version of Dong's proposed action for
the bulk and a cosmic brane with a certain tension \cite{Dong:2016fnf}, which
is applicable in the computation of holographic R\'{e}nyi entropies.

\section{Action on replica orbifold and cosmic branes\label{Section III}}

In the computation of holographic R\'{e}nyi entropies, it is useful to
consider the replica trick in order to construct a suitable $2n-$dimensional
bulk replica orbifold ${M}_{2n}^{\left(  \alpha\right)  }$, which is a
squashed cone (conically singular manifold without U$\left(  1\right)  $
isometry \cite{Fursaev:2013fta,atiyah_lebrun_2013}) having a conical angular
parameter $\alpha$, such that $2\pi\left(  1-\alpha\right)  $ is its angular
deficit. Then, using the AdS/CFT correspondence in the semi-classical limit,
the $m-th$ modular entropy $\widetilde{S}_{m}$
\cite{Dong:2016fnf,Nishioka:2018khk} which, as mentioned in the introduction,
is used in the computation of EREs, can be computed as%

\begin{equation}
\widetilde{S}_{m}=\left.  -\partial_{\alpha}I_{E}\left[  {M}_{2n}^{\left(
\alpha\right)  }\right]  \right\vert _{\alpha=\frac{1}{m}}, \label{SModular}%
\end{equation}
where $I_{E}\left[  {M}_{2n}^{\left(  \alpha\right)  }\right]  $ is the
Euclidean bulk gravitational action evaluated on ${M}_{2n}^{\left(
\alpha\right)  }$. As shown by Lewkowycz and Maldacena
\cite{Lewkowycz:2013nqa}, when one considers the Einstein-AdS action and the
limit of $\alpha\rightarrow1$, this prescription recovers the well-known RT
area formula \cite{Ryu:2006bv}, considering that the locus of the conical
sigularity (which is the fixed-point set of the replica symmetry) defines a
codimension-2 surface which coincides with $\Sigma$. In the case of
$\alpha=\frac{1}{m}$, with $m\in\mathbb{N}$ and $m>1$, the $m-th$ modular
entropy can be used to construct the $m-th$ R\'{e}nyi entropy $S_{m}$
according to Eq.(\ref{RenyiEnt}). In particular, as shown by Dong
\cite{Dong:2016fnf}, the locus of the conical singularity will correspond to
that of a cosmic brane with constant tension given by $T=\frac{\left(
1-\alpha\right)  }{4G}$, whose coupling to the bulk metric is implemented by
the Nambu-Goto (NG) action for the induced metric $\gamma$ on the brane.
Therefore, the conical defect of the replica orbifold ${M}_{2n}^{\left(
\alpha\right)  }$ is sourced by the cosmic brane and its location is
determined by minimizing the full action, which considers the contributions of
both the bulk Einstein-AdS action and the NG action of the cosmic brane. This
idea is further implemented by Nishioka \cite{Nishioka:2018khk}, where it is
explained that the resulting total action (which is refered to as the
bulk-per-replica action), contains the contribution at the conical
singularity, which precisely gives the usual area formula of RT and its
correct generalization beyond the tensionless limit, which is needed to
compute the modular entropy, and from it, the R\'{e}nyi entropy.

We first show how the evaluation of the standard Einstein-AdS action in the
replica orbifold ${M}_{2n}^{\left(  \alpha\right)  }$ directly leads to the
total action considered by Dong. Then, we consider its renormalized version,
in light of the volume renormalization procedure developed in Section
\ref{Section II}.

From the computation of curvature invariants defined on conically singular
manifolds
\cite{Fursaev:1995ef,Fursaev:2013fta,Mann:1996bi,Dahia:1998md,atiyah_lebrun_2013}%
, and in particular for the case of squashed-cone manifolds as studied by
Fursaev, Patrushev and Solodukhin in Ref.\cite{Fursaev:2013fta}, we have that%

\begin{equation}
R^{\left(  \alpha\right)  }=R+4\pi\left(  1-\alpha\right)  \delta_{\Sigma_{T}%
}, \label{RicciConical}%
\end{equation}
where $R^{\left(  \alpha\right)  }$ is the Ricci scalar of the orbifold
${M}_{2n}^{\left(  \alpha\right)  }$, $R$ is its regular part, $2\pi\left(
1-\alpha\right)  $ is the angular defect of the squashed cone and
$\delta_{\Sigma_{T}}$ is a $\left(  2n-2\right)  -$dimensional $\delta$
function which has support only at the location of the conical singularity
(which coincides with the on-shell position of the cosmic brane). Therefore,
by using the definition of the $\delta_{\Sigma_{T}}$, which is such that%

\begin{equation}%
%TCIMACRO{\dint \limits_{{M}_{2n}^{\left(  \alpha\right)  }}}%
%BeginExpansion
{\displaystyle\int\limits_{{M}_{2n}^{\left(  \alpha\right)  }}}
%EndExpansion
d^{2n}x\sqrt{G}\delta_{\Sigma_{T}}=%
%TCIMACRO{\dint \limits_{\Sigma_{T}}}%
%BeginExpansion
{\displaystyle\int\limits_{\Sigma_{T}}}
%EndExpansion
d^{2n-2}y\sqrt{\gamma},
\end{equation}
where $\Sigma_{T}$ is the codimension-2 geometric locus of the conical
singularity and $\sqrt{\gamma}$ is the induced metric at the $\Sigma_{T}$
surface, we have that%

\begin{equation}%
%TCIMACRO{\dint \limits_{{M}_{2n}^{\left(  \alpha\right)  }}}%
%BeginExpansion
{\displaystyle\int\limits_{{M}_{2n}^{\left(  \alpha\right)  }}}
%EndExpansion
d^{2n}x\sqrt{G}R^{\left(  \alpha\right)  }=%
%TCIMACRO{\dint \limits_{M_{2n}^{\left(  \alpha\right)  }\setminus\Sigma_{T}}}%
%BeginExpansion
{\displaystyle\int\limits_{M_{2n}^{\left(  \alpha\right)  }\setminus\Sigma
_{T}}}
%EndExpansion
d^{2n}x\sqrt{G}R+4\pi\left(  1-\alpha\right)
%TCIMACRO{\dint \limits_{\Sigma_{T}}}%
%BeginExpansion
{\displaystyle\int\limits_{\Sigma_{T}}}
%EndExpansion
d^{2n-2}y\sqrt{\gamma}.\label{RicciScalarConical}%
\end{equation}
Finally, considering that the NG action of a codimension-2 brane $\Sigma_{T}$
with tension T is given by%

\begin{equation}
I_{NG}\left[  \Sigma_{T}\right]  =T%
%TCIMACRO{\dint \limits_{\Sigma_{T}}}%
%BeginExpansion
{\displaystyle\int\limits_{\Sigma_{T}}}
%EndExpansion
d^{2n-2}y\sqrt{\gamma},
\end{equation}
we have that the Einstein-AdS action evaluated on ${M}_{2n}^{\left(
\alpha\right)  }$ gives%

\begin{equation}
I_{EH}\left[  {M}_{2n}^{\left(  \alpha\right)  }\right]  =\frac{1}{16\pi G}%
%TCIMACRO{\dint \limits_{M_{2n}^{\left(  \alpha\right)  }\setminus\Sigma_{T}}}%
%BeginExpansion
{\displaystyle\int\limits_{M_{2n}^{\left(  \alpha\right)  }\setminus\Sigma
_{T}}}
%EndExpansion
d^{2n}x\sqrt{G}\left(  R-2\Lambda\right)  +\frac{\left(  1-\alpha\right)
}{4G}%
%TCIMACRO{\dint \limits_{\Sigma_{T}}}%
%BeginExpansion
{\displaystyle\int\limits_{\Sigma_{T}}}
%EndExpansion
d^{2n-2}y\sqrt{\gamma},
\end{equation}
and therefore,%

\begin{align}
I_{EH}\left[  {M}_{2n}^{\left(  \alpha\right)  }\right]   &  =I_{EH}\left[
M_{2n}^{\left(  \alpha\right)  }\setminus\Sigma_{T}\right]  +I_{NG}\left[
\Sigma_{T}\right]  =I_{tot},\nonumber\\
T\left(  \alpha\right)   &  =\frac{\left(  1-\alpha\right)  }{4G},
\end{align}
which corresponds to the total action $I_{tot}$ considered by Dong. Hence, the
NG action arises as the conical contribution to the EH action evaluated on the
replica orbifold. The $m-th$ modular entropy $\widetilde{S}_{m}$ can then be
computed, according to Eq.(\ref{SModular}), thus obtaining Eq.(\ref{ModEnt}).
Finally, the holographic ERE can be computed according to Eq.(\ref{RenyiEnt}).
We now proceed to obtain the renormalized version of the total action
$I_{tot}^{ren}$, by considering the evaluation of the renormalized
Einstein-AdS action on the orbifold ${M}_{2n}^{\left(  \alpha\right)  }$.

\subsection{Renormalized NG action from the conical singularity}

We evaluate the renormalized Einstein-AdS action on the replica orbifold, obtaining%

\begin{equation}
I_{EH}^{ren}\left[  {M}_{2n}^{\left(  \alpha\right)  }\right]  =\frac{1}{16\pi
G}%
%TCIMACRO{\dint \limits_{{M}_{2n}^{\left(  \alpha\right)  }}}%
%BeginExpansion
{\displaystyle\int\limits_{{M}_{2n}^{\left(  \alpha\right)  }}}
%EndExpansion
d^{2n}x\sqrt{G}{\ell}^{2\left(  n-1\right)  }P_{2n}\left[  W_{\left(
E\right)  }^{\left(  \alpha\right)  }\right]  -\frac{c_{2n}}{16\pi G}\left(
4\pi\right)  ^{n}n!\chi\left[  {M}_{2n}^{\left(  \alpha\right)  }\right]  ,
\end{equation}
where the conically-singular Einstein Weyl is defined as%

\begin{equation}
\left.  W_{\left(  E\right)  }^{\left(  \alpha\right)  }\right.  _{\nu_{1}%
\nu_{2}}^{\mu_{1}\mu_{2}}\overset{\text{def}}{=}\left.  R^{\left(
\alpha\right)  }\right.  _{\nu_{1}\nu_{2}}^{\mu_{1}\mu_{2}}+\frac{1}{{\ell
}^{2}}\delta_{\left[  \nu_{1}\nu_{2}\right]  }^{\left[  \mu_{1}\mu_{2}\right]
},
\end{equation}
and $c_{2n}$ is given in Eq.(\ref{c_2n}). Now, using that%

\begin{equation}%
%TCIMACRO{\dint \limits_{{M}_{2n}^{\left(  \alpha\right)  }}}%
%BeginExpansion
{\displaystyle\int\limits_{{M}_{2n}^{\left(  \alpha\right)  }}}
%EndExpansion
d^{2n}x\sqrt{G}{\ell}^{2\left(  n-1\right)  }P_{2n}\left[  W_{\left(
E\right)  }^{\left(  \alpha\right)  }\right]  =%
%TCIMACRO{\dint \limits_{{M}_{2n}^{\left(  \alpha\right)  }}}%
%BeginExpansion
{\displaystyle\int\limits_{{M}_{2n}^{\left(  \alpha\right)  }}}
%EndExpansion
d^{2n}x\sqrt{G}\left(  R^{\left(  \alpha\right)  }-2\Lambda\right)  +c_{2n}%
%TCIMACRO{\dint \limits_{{M}_{2n}^{\left(  \alpha\right)  }}}%
%BeginExpansion
{\displaystyle\int\limits_{{M}_{2n}^{\left(  \alpha\right)  }}}
%EndExpansion
\mathcal{E}_{2n}^{\left(  \alpha\right)  },
\end{equation}
where $P_{2n}\left[  X\right]  $ is given in Eq.(\ref{P2n}) and $\varepsilon
_{2n}$ is defined in Eq.(\ref{EulerDensity}), we have%

\begin{equation}
I_{EH}^{ren}\left[  {M}_{2n}^{\left(  \alpha\right)  }\right]  =\frac{1}{16\pi
G}%
%TCIMACRO{\dint \limits_{{M}_{2n}^{\left(  \alpha\right)  }}}%
%BeginExpansion
{\displaystyle\int\limits_{{M}_{2n}^{\left(  \alpha\right)  }}}
%EndExpansion
d^{2n}x\sqrt{G}\left(  R^{\left(  \alpha\right)  }-2\Lambda\right)
+\frac{c_{2n}}{16\pi G}%
%TCIMACRO{\dint \limits_{{M}_{2n}^{\left(  \alpha\right)  }}}%
%BeginExpansion
{\displaystyle\int\limits_{{M}_{2n}^{\left(  \alpha\right)  }}}
%EndExpansion
\mathcal{E}_{2n}^{\left(  \alpha\right)  }-\frac{c_{2n}}{16\pi G}\left(
4\pi\right)  ^{n}n!\chi\left[  {M}_{2n}^{\left(  \alpha\right)  }\right]  .
\end{equation}
Also, from the properties of the Euler density for squashed cones as
conjectured in Ref.\cite{Anastasiou:2018rla}, we have that%

\begin{align}%
%TCIMACRO{\dint \limits_{{M}_{2n}^{\left(  \alpha\right)  }}}%
%BeginExpansion
{\displaystyle\int\limits_{{M}_{2n}^{\left(  \alpha\right)  }}}
%EndExpansion
\mathcal{E}_{2n}^{\left(  \alpha\right)  } &  =%
%TCIMACRO{\dint \limits_{M_{2n}^{\left(  \alpha\right)  }\setminus\Sigma_{T}}}%
%BeginExpansion
{\displaystyle\int\limits_{M_{2n}^{\left(  \alpha\right)  }\setminus\Sigma
_{T}}}
%EndExpansion
\mathcal{E}_{2n}+4\pi n\left(  1-\alpha\right)
%TCIMACRO{\dint \limits_{\Sigma_{T}}}%
%BeginExpansion
{\displaystyle\int\limits_{\Sigma_{T}}}
%EndExpansion
\varepsilon_{2n-2}~+O\left(  \left(  1-\alpha\right)  ^{2}\right)
,\nonumber\\
\chi\left[  {M}_{2n}^{\left(  \alpha\right)  }\right]   &  =\chi\left[
M_{2n}^{\left(  \alpha\right)  }\setminus\Sigma_{T}\right]  +\left(
1-\alpha\right)  \chi\left[  \Sigma_{T}\right]  +O\left(  \left(
1-\alpha\right)  ^{2}\right)  ,
\end{align}
and by considering Eq.(\ref{RicciScalarConical}), we can write $I_{EH}%
^{ren}\left[  {M}_{2n}^{\left(  \alpha\right)  }\right]  $ as%

\begin{align}
I_{EH}^{ren}\left[  {M}_{2n}^{\left(  \alpha\right)  }\right]   &  =\frac
{1}{16\pi G}\left(  \displaystyle\int\limits_{M_{2n}^{\left(  \alpha\right)
}\setminus\Sigma_{T}}d^{2n}x\sqrt{G}{\ell}^{2\left(  n-1\right)  }%
P_{2n}\left[  W_{\left(  E\right)  }\right]  -c_{2n}\left(  4\pi\right)
^{n}n!\chi\left[  M_{2n}^{\left(  \alpha\right)  }\setminus\Sigma_{T}\right]
\right)  \nonumber\\
+ &  \frac{\left(  1-\alpha\right)  }{4G}\left(  \displaystyle\int
\limits_{\Sigma_{T}}d^{2n-2}y\sqrt{\gamma}+nc_{2n}\displaystyle\int
\limits_{\Sigma_{T}}\varepsilon_{2n-2}-nc_{2n}\left(  4\pi\right)
^{n-1}\left(  n-1\right)  !\chi\left(  \Sigma_{T}\right)  \right)  \nonumber\\
&  +O\left(  \left(  1-\alpha\right)  ^{2}\right)  .
\end{align}
Finally using that (as shown in Ref.\cite{Anastasiou:2018rla})%

\begin{equation}%
%TCIMACRO{\dint \limits_{\Sigma_{T}}}%
%BeginExpansion
{\displaystyle\int\limits_{\Sigma_{T}}}
%EndExpansion
d^{2n-2}y\sqrt{\gamma}+nc_{2n}%
%TCIMACRO{\dint \limits_{\Sigma_{T}}}%
%BeginExpansion
{\displaystyle\int\limits_{\Sigma_{T}}}
%EndExpansion
\varepsilon_{2n-2}=-\frac{{\ell}^{2}}{2\left(  2n-3\right)  }%
%TCIMACRO{\dint \limits_{\Sigma_{T}}}%
%BeginExpansion
{\displaystyle\int\limits_{\Sigma_{T}}}
%EndExpansion
d^{2n-2}y\sqrt{\gamma}{\ell}^{2\left(  n-2\right)  }P_{2n-2}\left[
\mathcal{F}_{AdS}\right]  , \label{SigmaVolRenorm}%
\end{equation}
where $\mathcal{F}_{AdS}$ for $\Sigma_{T}$ is defined in Eq.(\ref{FAdS}), and that%

\begin{equation}
c_{2n-2}=-\frac{2\left(  2n-3\right)  }{{\ell}^{2}}nc_{2n},
\end{equation}
we obtain%

\begin{align}
I_{EH}^{ren}\left[  {M}_{2n}^{\left(  \alpha\right)  }\right]   &  =\frac
{1}{16\pi G}\displaystyle\int\limits_{M_{2n}^{\left(  \alpha\right)
}\setminus\Sigma_{T}}d^{2n}x\sqrt{G}{\ell}^{2\left(  n-1\right)  }%
P_{2n}\left[  W_{\left(  E\right)  }\right]  -\frac{c_{2n}}{16\pi G}\left(
4\pi\right)  ^{n}n!\chi\left[  M_{2n}^{\left(  \alpha\right)  }\setminus
\Sigma_{T}\right]  \nonumber\\
&  +\frac{\left(  1-\alpha\right)  }{4G}\left(  -\frac{{\ell}^{2}}{2\left(
2n-3\right)  }\right)  \Bigg(\displaystyle\int\limits_{\Sigma_{T}}%
d^{2n-2}y\sqrt{\gamma}{\ell}^{2\left(  n-2\right)  }P_{2n-2}\left[
\mathcal{F}_{AdS}\right]  \nonumber\\
&  -c_{2n-2}\left(  4\pi\right)  ^{n-1}\left(  n-1\right)  !\chi\left[
\Sigma_{T}\right]  \Bigg)+O\left(  \left(  1-\alpha\right)  ^{2}\right)  .
\end{align}
Therefore,%

\begin{equation}
I_{EH}^{ren}\left[  {M}_{2n}^{\left(  \alpha\right)  }\right]  =I_{EH}%
^{ren}\left[  M_{2n}^{\left(  \alpha\right)  }\setminus\Sigma_{T}\right]
+I_{NG}^{ren}\left[  \Sigma_{T}\right]  +O\left(  \left(  1-\alpha\right)
^{2}\right)  ,
\end{equation}
where $I_{EH}^{ren}\left[  M_{2n}^{\left(  \alpha\right)  }\setminus\Sigma
_{T}\right]  $ is given in Eq.(\ref{IRenPoly}), and $I_{NG}^{ren}\left[
\Sigma_{T}\right]  $ is defined as%

\begin{equation}
I_{NG}^{ren}\left[  \Sigma_{T}\right]  =\frac{\left(  1-\alpha\right)  }%
{4G}Vol_{ren}\left[  \Sigma_{T}\right]  , \label{IRenNG}%
\end{equation}
for $Vol_{ren}\left[  \Sigma_{T}\right]  $ given by%

\begin{equation}
Vol_{ren}\left[  \Sigma\right]  =-\frac{{\ell}^{2}}{2\left(  2n-3\right)
}\Bigg(
%TCIMACRO{\dint \limits_{\Sigma}}%
%BeginExpansion
{\displaystyle\int\limits_{\Sigma}}
%EndExpansion
d^{2n-2}y\sqrt{\gamma}{\ell}^{2\left(  n-2\right)  }P_{2n-2}\left[
\mathcal{F}_{AdS}\right]  -c_{2n-2}\left(  4\pi\right)  ^{n-1}\left(
n-1\right)  !\chi\left[  \Sigma\right]  \Bigg) , \label{VolRenSigma}%
\end{equation}
in complete analogy with the formula for the renormalized volume of the bulk
manifold. By defining%

\begin{equation}
I_{tot}^{ren}\overset{\text{def}}{=}I_{EH}^{ren}\left[  M_{2n}^{\left(
\alpha\right)  }\setminus\Sigma_{T}\right]  +I_{NG}^{ren}\left[  \Sigma
_{T}\right]  \text{,}\label{ITotalRenorm}%
\end{equation}
we note that our expression for $I_{tot}^{ren}$ is consistent with Dong's
proposal for $I_{tot}$, as it corresponds to the renormalized version of it.
Therefore, our renormalized total action is understood as the sum of the
renormalized Einstein-AdS action for the regular part of the bulk plus the
renormalized NG action of the cosmic brane. We emphasize that the dynamics
obtained by extremizing $I_{tot}^{ren}$ is the same as that obtained by
extremizing $I_{tot}$ as given in Ref.\cite{Dong:2016fnf}, because both
actions only differ by topological bulk terms that are equivalent to boundary
terms through use of the Euler theorem, and therefore, they lead to the same
bulk Euler-Lagrange equations of motion. Furthermore, the boundary terms
generated on the AdS boundary and on $\partial\Sigma_{T}$, are precisely the
ones that cancel the bulk divergences, while at the same time being consistent
with Dirichlet boundary conditions for the CFT metric $g_{\left(  0\right)  }$
and for the induced metric $\sigma_{\left(  0\right)  }$ on the entanglement
surface in the CFT, in the notation of
Ref.\cite{Anastasiou:2017xjr,Anastasiou:2018rla}.

We note that, $I_{EH}^{ren}\left[  {M}_{2n}^{\left(  \alpha\right)  }\right]
$ appears to differ from $I_{tot}^{ren}$ by an unspecified $O\left(  \left(
1-\alpha\right)  ^{2}\right)  $ part. However, we conjecture that there should
not be extra terms of higher order in $\left(  1-\alpha\right)  $ and, that
therefore, we should have%

\begin{align}%
%TCIMACRO{\dint \limits_{{M}_{2n}^{\left(  \alpha\right)  }}}%
%BeginExpansion
{\displaystyle\int\limits_{{M}_{2n}^{\left(  \alpha\right)  }}}
%EndExpansion
\mathcal{E}_{2n}^{\left(  \alpha\right)  } &  =%
%TCIMACRO{\dint \limits_{M_{2n}^{\left(  \alpha\right)  }\setminus\Sigma_{T}}}%
%BeginExpansion
{\displaystyle\int\limits_{M_{2n}^{\left(  \alpha\right)  }\setminus\Sigma
_{T}}}
%EndExpansion
\mathcal{E}_{2n}+4\pi n\left(  1-\alpha\right)
%TCIMACRO{\dint \limits_{\Sigma_{T}}}%
%BeginExpansion
{\displaystyle\int\limits_{\Sigma_{T}}}
%EndExpansion
\varepsilon_{2n-2},\nonumber\\
\chi\left[  {M}_{2n}^{\left(  \alpha\right)  }\right]   &  =\chi\left[
M_{2n}^{\left(  \alpha\right)  }\setminus\Sigma_{T}\right]  +\left(
1-\alpha\right)  \chi\left[  \Sigma_{T}\right]  ,\ \ \ \ \label{NewProposal}%
\end{align}
with no additional $O\left(  \left(  1-\alpha\right)  ^{2}\right)  $ pieces,
instead. The reasoning for this is that the location of the conical
singularity or, equivalently, the position of the brane, should not be changed
by the renormalization procedure. In turn, the bulk dynamics should not be
affected and, consequently, the extra bulk terms added to the NG action of
$\Sigma_{T}$ in order to achieve the renormalization should be equivalent to a
boundary term at $\partial\Sigma_{T}$, which is fixed by the boundary
conditions. This is precisely our case, as can be seen in
Eq.(\ref{SigmaVolRenorm}) for the $O\left(  \left(  1-\alpha\right)  \right)
$ part, as the $\varepsilon_{2n-2}$ is equivalent to the $\left(  n-1\right)
-$th Chern form $B_{2n-3}$ evaluated at $\partial\Sigma_{T}$ by virtue of the
Euler theorem in $\left(  2n-2\right)  $ dimensions. Hence, any extra
contribution located at $\delta_{\Sigma_{T}}$ of higher order in $\left(
1-\alpha\right)  $ would necessarily have to be an extra contribution
proportional to $\varepsilon_{2n-2}$, or to some other bulk topological term
dynamically equivalent to a boundary term at $\partial\Sigma_{T}$. In the
former case, the effect would be to change the form of the tension $T$ as a
function of $\alpha$, which is not allowed as it would change the physics.
Finally, the latter case seems very contrived and unlikely as other
topological terms are not always defined for $\left(  2n-2\right)
-$dimensional manifolds. Therefore, our conjecture of Eq.(\ref{NewProposal})
follows, on physical grounds, for the particular case of conical singularities
induced by cosmic branes with tension whose position is fixed by requiring the
on-shell condition. Finally, we then have that%

\begin{equation}
I_{EH}^{ren}\left[  {M}_{2n}^{\left(  \alpha\right)  }\right]  =I_{tot}^{ren},
\end{equation}
with $I_{tot}^{ren}$ as given in Eq.(\ref{ITotalRenorm}), in analogy with
Dong's total action for the unrenormalized case.

\subsection{Action on conical orbifold as deformed volume}

The renormalized volume of the conically singular manifold $M_{2n}^{(\alpha)}$
can be thought of as a one-parameter family of deformations with respect to
the renormalized volume of the non--singular manifold $M_{2n}^{\left(
\alpha\right)  }\setminus\Sigma_{T}$, that is
\begin{equation}
Vol_{ren}\big[M_{2n}^{(\alpha)}\big]=Vol_{ren}\big[M_{2n}^{\left(
\alpha\right)  }\setminus\Sigma_{T}\big]+q_{\alpha}~,\label{Vshift}%
\end{equation}
where $q_{\alpha}$ is a finite deformation that vanishes in the tensionless
limit of $\alpha\rightarrow1$. From the bulk perspective, Eq.\eqref{Vshift}
amounts to reinterpreting the problem of computing the renormalized modular
entropy $\widetilde{S}_{m}^{ren}$, which in turns yields a renormalized
R\'{e}nyi entropy, as the problem of extremizing a bulk hypersurface while
fixing the renormalized $q$--deformed volume bounded by it.

Starting from $Vol_{ren}\left[  M_{2n}^{(\alpha)}\right]  $ as given in
Eq.(\ref{VolRenM2n}) and considering that, as discussed above, $I_{EH}%
^{ren}\left[  {M}_{2n}^{\left(  \alpha\right)  }\right]  =I_{tot}^{ren}$, for
$I_{tot}^{ren}$ given in Eq.(\ref{ITotalRenorm}), we can write the
renormalized volume of the replica orbifold as
\begin{equation}
Vol_{ren}\big[M_{2n}^{(\alpha)}\big]=Vol_{ren}\big[M_{2n}^{(\alpha)}%
\setminus\Sigma_{T}\big]-\left(  1-\alpha\right)  \frac{2\pi{\ell}^{2}%
}{\left(  2n-1\right)  }Vol_{ren}\left[  \Sigma_{T}\right]
.\label{Shifted_Bulk_Volume}%
\end{equation}
Therefore, the volume deformation $q_{\alpha}$ is given by%

\begin{equation}
q_{\alpha}=-\left(  1-\alpha\right)  \frac{2\pi{\ell}^{2}}{\left(
2n-1\right)  }Vol_{ren}\left[  \Sigma_{T}\right]  , \label{Shift}%
\end{equation}
and it is labelled by the angular parameter $\alpha$, or equivalently by the
tension $T\left(  \alpha\right)  =\frac{\left(  1-\alpha\right)  }{4G}$. This
deformation has support only at the codimension-2 surface $\Sigma_{T}$.

As an example, we consider the $n=2$ case for an AAdS$_{4}$ bulk, where the
renormalized volume evaluated on the orbifold becomes%

\begin{align}
Vol_{ren}\left[  M_{4}^{\left(  \alpha\right)  }\right]   &  =-\frac{{\ell
}^{4}}{96}%
%TCIMACRO{\dint \limits_{M_{4}}}%
%BeginExpansion
{\displaystyle\int\limits_{M_{4}}}
%EndExpansion
d^{4}x\sqrt{G}\delta_{\left[  \mu_{1}\mu_{2}\mu_{3}\mu_{4}\right]  }^{\left[
\nu_{1}\nu_{2}\nu_{3}\nu_{4}\right]  }\left.  W_{\left(  E\right)  }\right.
_{\nu_{1}\nu_{2}}^{\mu_{1}\mu_{2}}\left.  W_{\left(  E\right)  }\right.
_{\nu_{3}\nu_{4}}^{\mu_{3}\mu_{4}}+\frac{4\pi^{2}{\ell}^{4}}{3}\chi\left[
M_{4}\right]  \nonumber\\
&  -\left(  1-\alpha\right)  \left(  \frac{\pi{\ell}^{4}}{6}%
%TCIMACRO{\dint \limits_{\Sigma_{T}}}%
%BeginExpansion
{\displaystyle\int\limits_{\Sigma_{T}}}
%EndExpansion
d^{2}y\sqrt{\gamma}\delta_{\left[  a_{1}a_{2}\right]  }^{\left[  b_{1}%
b_{2}\right]  }\left.  \mathcal{F}_{AdS}\right.  _{b_{1}b_{2}}^{a_{1}a_{2}%
}-\frac{4\pi^{2}{\ell}^{4}}{3}\chi\left[  \Sigma_{T}\right]  \right)
\end{align}

and therefore, in this case, the deformation in the bulk volume is given by%

\begin{equation}
q_{\alpha}=-\left(  1-\alpha\right)  \left(  \frac{\pi{\ell}^{4}}{6}%
%TCIMACRO{\dint \limits_{\Sigma_{T}}}%
%BeginExpansion
{\displaystyle\int\limits_{\Sigma_{T}}}
%EndExpansion
d^{2}y\sqrt{\gamma}\delta_{\left[  a_{1}a_{2}\right]  }^{\left[  b_{1}%
b_{2}\right]  }\left.  \mathcal{F}_{AdS}\right.  _{b_{1}b_{2}}^{a_{1}a_{2}%
}-\frac{4\pi^{2}{\ell}^{4}}{3}\chi\left[  \Sigma_{T}\right]  \right)  .
\end{equation}

\section{Renormalized R\'{e}nyi entropy from renormalized area of cosmic
branes\label{Section IV}}

Considering $I_{tot}^{ren}$ as defined in Eq.(\ref{ITotalRenorm}), the
renormalized modular entropy can be trivially computed as%

\begin{equation}
\widetilde{S}_{m}^{ren}=-\partial_{\alpha}I_{tot}^{ren}=-\partial_{\alpha
}I_{NG}^{ren},
\end{equation}
in accordance with the discussion given in the introduction. Therefore, using
the form of $I_{NG}^{ren}$ defined in Eq.(\ref{IRenNG}), we obtain%

\begin{equation}
\widetilde{S}_{m}^{ren}=\frac{Vol_{ren}\left[  \Sigma_{T}\right]  }{4G},
\end{equation}
where $T\left(  m\right)  =\frac{\left(  m-1\right)  }{4mG}$, in agreement
with Eq.(\ref{ModEnt}). Now, the renormalized ERE can be computed from
$\widetilde{S}_{m}^{ren}$ using Eq.(\ref{RenyiEnt}), such that%

\begin{equation}
S_{m}^{ren}=\frac{m}{m-1}%
%TCIMACRO{\dint \limits_{1}^{m}}%
%BeginExpansion
{\displaystyle\int\limits_{1}^{m}}
%EndExpansion
\frac{dm^{\prime}}{{m^{\prime}}^{2}}\widetilde{S}_{{m}^{\prime}}^{ren}.
\end{equation}

We remark that in computing $\partial_{\alpha}I_{tot}^{ren}$, the $\alpha$
dependence of the location of $\Sigma_{T}$ should no be taken into account,
because first one requires its location to be determined by the extremization
of $I_{tot}^{ren}$, and then, this location is taken as a given.

Now, for illustrative purposes, we present the form adopted by the modular
entropy for the cases of AAdS$_{4}$ and AAdS$_{6}$ bulk manifolds.

\subsection{Modular entropy in AdS$_{4}$/CFT$_{3}$ and in AdS$_{6}$/CFT$_{5}$}

For an AAdS$_{4}$ bulk manifold and an embedded codimension-2 surface
$\Sigma_{T}$ whose position is determined by minimizing $I_{tot}^{ren}$ for a
cosmic brane with tension $T\left(  m\right)  $, such that $\partial\Sigma
_{T}$ lies at the AdS boundary, we have that the modular entropy in the
CFT$_{3}$ is given by%

\begin{equation}
\widetilde{S}_{m}^{ren}=-\partial_{\alpha}I_{tot}^{ren}=\frac{{\ell}^{2}}{16G}%
%TCIMACRO{\dint \limits_{\Sigma_{T}}}%
%BeginExpansion
{\displaystyle\int\limits_{\Sigma_{T}}}
%EndExpansion
d^{2}y\sqrt{\gamma}\delta_{\left[  a_{1}a_{2}\right]  }^{\left[  b_{1}%
b_{2}\right]  }\left.  \mathcal{F}_{AdS}\right.  _{b_{1}b_{2}}^{a_{1}a_{2}%
}-\frac{\pi{\ell}^{2}}{2G}\chi\left[  \Sigma_{T}\right]  , \label{SModRenn2}%
\end{equation}
where $\mathcal{F}_{AdS}$ is defined in Eq.(\ref{FAdS}), and $\chi\left[
\Sigma_{T}\right]  $ is the Euler characteristic of $\Sigma_{T}$. We note that
in the tensionless limit, we recover our result for $S_{EE}^{ren}$ given in
Ref.\cite{Anastasiou:2017xjr}. Also, the resulting expression for
$\widetilde{S}_{m}^{ren}$, when multiplied by $4G$, matches the formula given
by Alexakis and Mazzeo for the renormalized area of a codimension-2 minimal
surface embedded in a four-dimensional AAdS Einstein manifold, given in
Ref.\cite{Alexakis2010}.

Analogously, for an AAdS$_{6}$ bulk manifold, the corresponding renormalized
modular entropy of the CFT$_{5}$ is given by%

\begin{align}
\widetilde{S}_{m}^{ren}  &  =-\frac{{\ell}^{2}}{48G}\displaystyle\int
\limits_{\Sigma_{T}}d^{4}y\sqrt{\gamma}\left(  \frac{{\ell}^{2}}{8}%
\delta_{\left[  a_{1}a_{2}a_{3}a_{4}\right]  }^{\left[  b_{1}b_{2}b_{3}%
b_{4}\right]  }\left.  \mathcal{F}_{AdS}\right.  _{b_{1}b_{2}}^{a_{1}a_{2}%
}\left.  \mathcal{F}_{AdS}\right.  _{b_{3}b_{4}}^{a_{3}a_{4}}\right.
\nonumber\\
&  -\left.  \delta_{\left[  a_{1}a_{2}\right]  }^{\left[  b_{1}b_{2}\right]
}\left.  \mathcal{F}_{AdS}\right.  _{b_{1}b_{2}}^{a_{1}a_{2}}\right)
+\frac{\pi^{2}{\ell}^{4}}{3G}\chi\left[  \Sigma_{T}\right]  .
\label{SModRenn3}%
\end{align}
We note that in the tensionless limit, we recover our result for $S_{EE}%
^{ren}$ in the AdS$_{6}$/CFT$_{5}$ case, as given in
Ref.\cite{Anastasiou:2018rla}. We also note that, when multiplying by $4G$,
the resulting expression is equal to the renormalized volume of a
four-dimensional AAdS manifold, having the same structure as the renormalized
Einstein-AdS$_{4}$ action, but with an extra $\delta_{\left[  a_{1}%
a_{2}\right]  }^{\left[  b_{1}b_{2}\right]  }\left.  \mathcal{F}_{AdS}\right.
_{b_{1}b_{2}}^{a_{1}a_{2}}$ term, which is equal to $2tr\left[  \mathcal{F}%
_{AdS}\right]  $. For Einstein manifolds, the Weyl tensor $W_{\left(
E\right)  }$ is equal to $\mathcal{F}_{AdS}$, and because $tr\left[
W_{\left(  E\right)  }\right]  =0$, this term vanishes. This shows that
although the minimal surface $\Sigma_{T}$ is not an Einstein manifold, its
renormalized volume has the same form. The reason for this is that our
renormalized volume formula given in Eq.(\ref{VolRenM2n}), evaluated using
$\mathcal{F}_{AdS}$, is valid for AAdS Einstein manifolds and also for minimal
submanifolds of codimension-2 embedded therein, where the meaning of minimal
is that the submanifold extremizes the $I_{tot}^{ren}$ action.

\section{Example: Renormalized R\'{e}nyi entropies for ball-shaped entangling
regions in odd-dimensional CFTs\label{Section V}}

We now compute the renormalized ERE in the case of a ball-shaped entangling
region on the CFT. This case is of interest as it can be computed exactly, and
in the limit of $m\rightarrow1$ (tensionless limit), one can directly check
that the result obtained for $S_{EE}^{ren}$ in Ref.\cite{Anastasiou:2018rla}
is recovered.

The first computation of EREs for ball-shaped regions in holographic CFTs was
performed by Hung, Myers, Smolkin and Yale in Ref.\cite{Hung:2011nu}, by using
the Casini-Huerta-Myers (CHM) map \cite{Casini:2011kv} to relate the
computation of the modular entropy (used to obtain the ERE) to that of the
Bekenstein-Hawking entropy of a certain topological BH. In particular, by
considering a conformal transformation, the CFT was put in a hyperbolic
background geometry, such that the reduced density matrix for the entangling
region in the vacuum state was unitarily mapped into the thermal density
matrix of a Gibbs state. By considering the usual holographic embedding of the
CFT into global AdS, the conformal transformation in the CFT was seen to
correspond to a coordinate transformation in the bulk, mapping two different
foliations of AdS space. Also, by the standard AdS/CFT dictionary, the thermal
state in the CFT was identified as the holographic dual of the topological
black hole, whose (Euclidean) metric is given by%

\begin{equation}
ds^{2}=\frac{{\ell}^{2}}{R^{2}}\left[  f\left(  r\right)  d\tau^{2}%
+\frac{dr^{2}}{f\left(  r\right)  }+r^{2}\left(  du^{2}+\sinh^{2}%
u~d\Omega_{d-2}^{2}\right)  \right]  ,
\end{equation}
where%

\begin{equation}
f\left(  r\right)  =\frac{r^{2}}{R^{2}}-1-\frac{r_{H}^{d-2}}{r^{d-2}}\left(
\frac{r_{H}}{R^{2}}-1\right)  ,
\end{equation}
$d\Omega_{d-2}^{2}$ is the line element in a unit $\left(  d-2\right)  -$
sphere, $R$ is the radius of the original ball-shaped entangling region,
$r_{H}$ is the horizon radius where $f\left(  r\right)  $ vanishes and
$d=2n-1$ is the dimension of the CFT. Following Ref.\cite{Nishioka:2018khk},
from the metric we see that the topological BH has a temperature given by%

\begin{equation}
T\left(  x\right)  =\frac{1}{4\pi R}\left[  dx-\frac{d-2}{x}\right]  ,
\label{Tgeom}%
\end{equation}
where $x=\frac{r_{H}}{R}$ is the dimensionless horizon radius. Now, by
considering the form of the density matrix in the Gibbs state, the temperature
should also be equal to%

\begin{equation}
T_{m}=\frac{1}{2\pi Rm}, \label{TGibbs}%
\end{equation}
where $m$ is the replica index (such that we are computing the $m-th$ modular
entropy). By equating the expressions given in Eqs.(\ref{Tgeom}) and
(\ref{TGibbs}), we can solve for the dimensionless horizon radius $x_{m}$,
obtaining that%

\begin{equation}
x_{m}=\frac{1+\sqrt{dm^{2}\left(  d-2\right)  +1}}{dm}. \label{x_m}%
\end{equation}
Therefore, the Bekenstein-Hawking entropy of the BH, and consequently (through
the CHM map) the $m-th$ modular entropy in the CFT$_{d}$, is given by%

\begin{equation}
\widetilde{S}_{m}=\frac{Vol\left[  \mathcal{H}^{d-1}\right]  {\ell}^{d-1}}%
{4G}x_{m}^{d-1}, \label{ModSph}%
\end{equation}
where $\mathcal{H}^{d-1}$ denotes a constant curvature $\left(  d-1\right)
-$dimensional hyperbolic space with unit AdS radius. Then, finally, using
Eq.(\ref{RenyiEnt}), the $m-th$ R\'{e}nyi entropy is given by%

\begin{equation}
S_{m}=\frac{m}{m-1}%
%TCIMACRO{\dint \limits_{1}^{m}}%
%BeginExpansion
{\displaystyle\int\limits_{1}^{m}}
%EndExpansion
\frac{d{m}^{\prime}}{{m^{\prime}}^{2}}\widetilde{S}_{{m}^{\prime}}%
=\frac{Vol\left[  \mathcal{H}^{d-1}\right]  {\ell}^{d-1}}{4G}\frac{m}{2\left(
m-1\right)  }\left(  2-x_{m}^{d-2}-x_{m}^{d}\right)  . \label{RenyiSph}%
\end{equation}

We note that $Vol\left[  \mathcal{H}^{d-1}\right]  $ appearing in both the
modular entropy given in Eq.(\ref{ModSph}) and in the ERE given in
Eq.(\ref{RenyiSph}) is infinite, and therefore, the corresponding entropies
are divergent. We now proceed to renormalize them considering that
$\mathcal{H}^{d-1}$ is a $\left(  2n-2\right)  -$dimensional AAdS Einstein
manifold (in particular, it is global AdS), and therefore, the correctly
renormalized $Vol_{ren}\left[  \mathcal{H}^{d-1}\right]  $ can be computed
using our formula as given in Eq.(\ref{VolRenSigma}). Considering that
$\mathcal{H}^{d-1}$ has constant Riemannian curvature, and therefore
$\mathcal{F}_{\left.  AdS\right\vert _{\mathcal{H}^{d-1}}}=0$, and also that
its Euler characteristic $\chi\left[  \mathcal{H}^{d-1}\right]  =1$, we obtain that%

\begin{equation}
Vol_{ren}\left[  \mathcal{H}^{d-1}\right]  =\frac{c_{d-1}\left(  4\pi\right)
^{\frac{d-1}{2}}\left(  \frac{d-1}{2}\right)  !}{2\left(  d-2\right)  },
\end{equation}
and using that%

\begin{equation}
c_{d-1}=\frac{2\left(  -1\right)  ^{\frac{d-1}{2}}}{\left(  d-1\right)
\left(  d-3\right)  !},
\end{equation}
we have that%

\begin{equation}
Vol_{ren}\left[  \mathcal{H}^{d-1}\right]  =\frac{\left(  -1\right)
^{\frac{d-1}{2}}\left(  4\pi\right)  ^{\frac{d-1}{2}}\left(  \frac{d-1}%
{2}\right)  !}{\left(  d-1\right)  !}.
\end{equation}
We therefore find that the renormalized modular entropy and renormalized ERE
are given by%

\begin{equation}
\widetilde{S}_{m}^{ren}=\frac{\left(  -1\right)  ^{\frac{d-1}{2}}\left(
4\pi\right)  ^{\frac{d-1}{2}}\left(  \frac{d-1}{2}\right)  !{\ell}^{d-1}%
}{4G\left(  d-1\right)  !}x_{m}^{d-1}%
\end{equation}
and%

\begin{equation}
S_{m}^{ren}=\frac{\left(  -1\right)  ^{\frac{d-1}{2}}\left(  4\pi\right)
^{\frac{d-1}{2}}\left(  \frac{d-1}{2}\right)  !{\ell}^{d-1}}{4G\left(
d-1\right)  !}\frac{m}{2\left(  m-1\right)  }\left(  2-x_{m}^{d-2}-x_{m}%
^{d}\right)  .
\end{equation}

We now check that in the $m\rightarrow1$ limit, we recover the standard result
for the universal part of the entanglement entropy obtained by Kawano,
Nakaguchi and Nishioka in Ref.\cite{Kawano:2014moa}, which is equal to
$S_{EE}^{ren}$ as discussed in Ref.\cite{Anastasiou:2018rla}. Considering
Eq.(\ref{x_m}), we evaluate%

\begin{equation}
\lim\limits_{m\rightarrow1}x_{m}=1
\end{equation}
and%

\begin{equation}
\lim\limits_{m\rightarrow1}\frac{m}{2\left(  m-1\right)  }\left(
2-x_{m}^{d-2}-x_{m}^{d}\right)  =1,
\end{equation}
and therefore we obtain that%

\begin{equation}
\lim\limits_{m\rightarrow1}\widetilde{S}_{m}^{ren}=\lim\limits_{m\rightarrow
1}S_{m}^{ren}=\frac{\left(  -1\right)  ^{\frac{d-1}{2}}\left(  4\pi\right)
^{\frac{d-1}{2}}\left(  \frac{d-1}{2}\right)  !{\ell}^{d-1}}{4G\left(
d-1\right)  !},
\end{equation}
which precisely corresponds to $S_{EE}^{ren}$ as it should.

We comment more on the utility of computing the renormalized ERE, and on the
physical applications of our renormalized volume formula, in the next section.

\section{Outlook\label{Section VI}}

We conjectured the proportionality between renormalized Einstein-AdS action
and renormalized bulk volume for even-dimensional AAdS Einstein manifolds (see
Section \ref{Section II}). We also compared our proposal for the renormalized
volume with existing expressions given in the conformal geometry literature
for the four-dimensional \cite{2000math.....11051A} and six-dimensional
\cite{Yang2008} cases. Our resulting formula for the renormalized volume,
which applies for $2n-$dimensional AAdS Einstein manifolds, is given in
Eq.(\ref{VolRenM2n}) and has the form of a polynomial in full contractions of
powers of the Weyl tensor plus a topological constant that depends on the
Euler characteristic of the manifold. Therefore, our formula corresponds to a
concrete realization of Albin's prescription given in Ref.\cite{ALBIN2009140}.

We also reinterpreted modular entropies (which are used in the computation of
EREs) in terms of the renormalized areas of codimension-2 cosmic branes
$\Sigma_{T}$ with tension (see Section \ref{Section IV}), whose border
$\partial\Sigma_{T}$ is situated at the AdS boundary (being conformally
cobordant with the entangling region in the CFT), and whose precise location
is determined by requiring the total configuration to be a minimum of the
$I_{tot}^{ren}$ action as given in Eq.(\ref{ITotalRenorm}), which in turn is
the renormalized version of Dong's total action prescription
\cite{Dong:2016fnf} including the NG action of the brane. The obtained formula
for the renormalized volume of $\Sigma_{T}$ given in Eq.(\ref{VolRenSigma}),
has the same form as the renormalized volume for the bulk manifold, but in
codimension-2, and therefore, the same factors in the polynomial expansion are
recovered by replacing $n$ for $\left(  n-1\right)  $. However, there is one
important difference: in the case of $Vol_{ren}\left[  \Sigma_{T}\right]  $,
the polynomial is evaluated on $\mathcal{F}_{AdS}$ (as given in Eq.(\ref{FAdS}%
)) instead of on the Weyl tensor. Of course, in the case of Einstein
manifolds, both tensors are the same, but because the minimal codimension-2
surfaces $\Sigma_{T}$ (minimal in the sense of minimizing $I_{tot}^{ren}$)
need not be an Einstein manifold, the renormalized volume formula is seen to
be more general. Thus, we propose that the renormalized volume formula of
Eq.(\ref{VolRenM2n}) is valid for both AAdS Einstein manifolds and
codimension-2 minimal submanifolds embedded therein, such that $D=2n$ is the
dimension of the corresponding manifold or submanifold, where in general, the
polynomial is evaluated on $\mathcal{F}_{AdS}$. Our expression for the
renormalized volume can be understood as a measure of the deviation of a
manifold from the maximally symmetric constant curvature case, for which it is
only given by a constant proportional to the Euler characteristic of the
manifold. The obtained formula for the renormalized area of $\Sigma_{T}$
agrees with the formula given by Alexakis and Mazzeo in
Ref.\cite{Alexakis2010}, for the case when the brane is a 2-dimensional
extremal surface embedded in AAdS$_{4}$.

We also explicitated the relation between the geometrical interpretation of
the Einstein-AdS action evaluated on the conically singular replica orbifold,
as a one-parameter family of deformations to the renormalized bulk volume, and
Dong's codimension-2 cosmic brane construction of the total action
\cite{Dong:2016fnf} which includes the NG action of the brane, but in
renormalized form. We showed (in Section \ref{Section III}) that both
approaches are equivalent, and that the contribution to the bulk action at the
codimension-2 locus of the conical singularity is precisely the renormalized
NG action of the cosmic brane with a tension given by $T=\frac{\left(
m-1\right)  }{4mG}$, where $m$ is the replica index.

For the case of ball-shaped entangling regions (see Section \ref{Section V}),
we computed the renormalized ERE in $\left(  2n-1\right)  -$dimensional
holographic CFTs, following the computation performed by Hung, Myers, Smolkin
and Yale \cite{Hung:2011nu} using the CHM map \cite{Casini:2011kv}, and
renormalizing the horizon area of the corresponding topological BH. We have
also explicitly checked that in the tensionless limit, the known results for
the renormalized EE \cite{Anastasiou:2018rla} are correctly reproduced. This
case is of interest because, as discussed in Ref.\cite{Anastasiou:2018rla}, in
the tensionless limit, the renormalized EE is directly related to the $a_{d}%
-$charge \cite{Myers:2010xs}, which is a $C-$function candidate quantity that
decreases along renormalization group (RG) flows between conformal fixed
points and it counts the number of degrees of freedom of the CFT, providing a
generalization of the $c-$theorem \cite{Zamolodchikov:1986gt}.

The equivalence between the Kounterterms-renormalized Einstein-AdS action and
the renormalized volume is of interest as it recasts the problem of action
renormalization in gravity into the framework of volume renormalization in
conformal geometry. It therefore constitutes a mathematical validation of the
Kounterterms scheme, and it also emphasizes the topological nature of the
renormalized action, which has been systematically overlooked by the standard
Holographic Renormalization framework. For the recent discussions about
holographic complexity
\cite{Alishahiha:2015rta,Stanford:2014jda,Brown:2015bva,Abt:2017pmf,Banerjee:2017qti}%
, this result is interesting as it suggests a relation between the
\textit{complexity equals action} (CA) \cite{Brown:2015bva} and the
\textit{complexity equals volume} (CV) \cite{Stanford:2014jda} proposals. We
note, however, that our result does not imply that both proposals are directly
equivalent, as the volume considered in the CV proposal is an extremal
codimension-1 volume at constant boundary time, while the renormalized action
considered in the CA proposal is integrated over a region of the full
spacetime manifold. In trying to relate the two proposals, there may also be
more subtleties regarding the different domains of integration considered in
both (e.g., the Wheeler-de Witt patch in the CA case and the extremal spatial
slice crossing the Einstein-Rosen bridge in the CV case, when computing the
complexity of the thermofield-double state), and the differences between
Lorentzian and Euclidean gravity, which nonetheless are beyond the scope of
this paper.

The new computational scheme for obtaining renormalized EREs from the
renormalized volumes of codimension-2 minimal cosmic branes is interesting,
because as discussed by Headrick in Ref.\cite{Headrick:2010zt}, the EREs
encode the information of the full eigenvalue spectrum of the reduced density
matrix for the entangling region in the CFT, which has potential applications
for state reconstruction or, proceeding in reverse, for bulk geometry
reconstruction starting from the CFT. Furthermore, as discussed by Hung,
Myers, Smolkin and Yale in Ref.\cite{Hung:2011nu}, the EREs are, in general,
non-linear functions of the central charges and other CFT parameters, and
therefore, they are useful for characterizing CFTs and their behavior, for
example under RG flows, providing extra tools for a more detailed analysis
than what is possible from the renormalized EE only.

As future work, we will examine the significance of the equivalence between
renormalized action and renormalized volume for the study of holographic
complexity and its corresponding renormalization, revisiting the hinted
equivalence between the CA and CV proposals for the case of Einstein-AdS
gravity. We will also analyze the issue of volume renormalization for
odd-dimensional AAdS Einstein manifolds, attempting to relate it to the
Kounterterms-renormalized Einstein-AdS$_{2n+1}$ action presented in
Ref.\cite{Olea:2006vd}. Finally, we will investigate possible additions to the
holographic dictionary by considering a bulk configuration with a series of
embedded branes of different codimension, such that the full configuration is
required to be the minimum of an extended total action (in the spirit of
$I_{tot}^{ren}$ as given in Eq.(\ref{ITotalRenorm})), including the
corresponding codimension-$k$ renormalized NG actions for the new objects.
This renormalized action, constructed as a sum over different renormalized
volumes of objects with different codimension, should correspond to a
generalized notion of complexity \cite{Carmi:2017ezk} which seems worthy of
further enquiry.

\begin{acknowledgments}
The authors thank D.E. D\'{i}az, P. Sundell and A. Waldron for interesting discussions. C.A. is a Universidad Andres Bello (UNAB) Ph.D. Scholarship holder, and
his work is supported by Direcci\'{o}n General de Investigaci\'{o}n
(DGI-UNAB). This work is funded in part by FONDECYT Grants No. 1170765 ``\textit{Boundary dynamics in anti-de Sitter gravity and gauge/gravity duality}'' and No.
3180620 ``\textit{Entanglement Entropy and AdS gravity}'', and CONICYT Grant DPI 20140115.
\end{acknowledgments}\appendix

\section{Six-dimensional conformal invariants\label{Appendix A}}

In Eq.(\ref{IdentityOsborn}) of Section \ref{Section II}, we made use of an
identity given by Osborn and Stergiou in Ref.\cite{Osborn:2015rna}, for the
particular case of AAdS Einstein manifolds in six dimensions. The general case
of the identity, which according to the authors applies for
arbitrary-dimensional manifolds that need not be Einstein, states that%

\begin{align}
4I_{1}\left[  W\right]  -I_{2}\left[  W\right]   &  =W^{\rho\mu\nu\lambda
}\square W_{\rho\mu\nu\lambda}-2\left(  d-2\right)  S_{\rho}^{\sigma}%
W^{\rho\mu\nu\lambda}W_{\sigma\mu\nu\lambda}-2SW^{\rho\mu\nu\lambda}W_{\rho
\mu\nu\lambda}+\nonumber\\
&  2\left(  d-2\right)  \left(  d-3\right)  C^{\mu\nu\lambda}C_{\mu\nu\lambda
}+2\left(  d-2\right)  \nabla_{\sigma}\left(  W^{\sigma\mu\nu\lambda}C_{\mu
\nu\lambda}\right)  ,
\end{align}
where $W$ is the Weyl tensor, $I_{1}$ and $I_{2}$ are given in
Eq.(\ref{WeylInvs}), $S_{\rho}^{\sigma}$ is the Schouten tensor, $C^{\mu
\nu\lambda}$ is the Cotton tensor, $S$ is the trace of the Schouten tensor,
$\square\overset{\text{def}}{=}\nabla_{\mu}\nabla^{\mu}$ is the covariant
Laplacian operator and $d$ is the dimension of the bulk spacetime. Now, in the
case under study, for $d=6$ and considering an AAdS Einstein bulk manifold, we
have that%

\begin{equation}%
\begin{tabular}
[c]{l}%
$W_{\mu_{1}\mu_{2}}^{\nu_{1}\nu_{2}}=\left.  W_{\left(  E\right)  }\right.
_{\mu_{1}\mu_{2}}^{\nu_{1}\nu_{2}},$\\
$C_{\mu\nu\lambda}=0,$\\
$S=\frac{R}{2\left(  d-1\right)  }=\frac{-d\left(  d-1\right)  }{2\left(
d-1\right)  }=-3,$\\
$S_{\rho}^{\sigma}=-\frac{1}{2}\delta_{\rho}^{\sigma}.$%
\end{tabular}
\end{equation}
Therefore, we have that%

\begin{equation}
4I_{1}-I_{2}=\left.  W_{\left(  E\right)  }\right.  ^{\rho\mu\nu\lambda
}\square\left.  W_{\left(  E\right)  }\right.  _{\rho\mu\nu\lambda
}+4\left\vert W_{\left(  E\right)  }\right\vert ^{2}+6\left\vert W_{\left(
E\right)  }\right\vert ^{2},
\end{equation}
thus recovering Eq.(\ref{IdentityOsborn}) as given in the main body of the text.

\section{Vanishing of the neglected boundary term\label{Appendix B}}

The boundary term referred to as b.t. in Eq.(\ref{BTEq}), when written as a
covariant bulk contribution, is given by%

\begin{equation}
\text{b.t.}=\nabla_{\alpha}\left(  W_{\mu\nu}^{\beta\gamma}\nabla^{\alpha
}W_{\beta\gamma}^{\mu\nu}\right)  ,
\end{equation}
where $\nabla_{\alpha}$ is the bulk covariant derivative and $W_{\gamma\mu\nu
}^{\beta}$ is the bulk Weyl tensor. In the discussion of Section \ref{2.2} we
claim that this term vanishes asymptotically near the AdS boundary and we
therefore neglect it. We now proceed to show this by explicit computation.
Considering the radial foliation along the holographic coordinate $\rho$ with
the corresponding FG asymptotic expansions as given in
Refs.\cite{Anastasiou:2017xjr,Anastasiou:2018rla}, and considering that the
boundary of the manifold is the AdS boundary located at $\rho=0$, we have that%

\begin{align}%
%TCIMACRO{\dint \limits_{M_{6}}}%
%BeginExpansion
{\displaystyle\int\limits_{M_{6}}}
%EndExpansion
d^{6}x\sqrt{G}\nabla_{\alpha}\left(  W_{\mu\nu}^{\beta\gamma}\nabla^{\alpha
}W_{\beta\gamma}^{\mu\nu}\right)   &  =\lim\limits_{\varepsilon\rightarrow0}%
%TCIMACRO{\dint \limits_{\left.  \partial M_{6}\right\vert _{\rho=\varepsilon
%}}}%
%BeginExpansion
{\displaystyle\int\limits_{\left.  \partial M_{6}\right\vert _{\rho
=\varepsilon}}}
%EndExpansion
d^{5}x\sqrt{h}\left(  W_{\mu\nu}^{\beta\gamma}\nabla^{\rho}W_{\beta\gamma
}^{\mu\nu}\right) \nonumber\\
&  =\lim\limits_{\varepsilon\rightarrow0}\frac{1}{2}%
%TCIMACRO{\dint \limits_{\left.  \partial M_{6}\right\vert _{\rho=\varepsilon
%}}}%
%BeginExpansion
{\displaystyle\int\limits_{\left.  \partial M_{6}\right\vert _{\rho
=\varepsilon}}}
%EndExpansion
d^{5}x\sqrt{h}G^{\rho\rho}\partial_{\rho}\left(  \left\vert W\right\vert
^{2}\right)  .
\end{align}
Now, both the square of the bulk Weyl tensor $\left\vert W\right\vert ^{2}$
and $G^{\rho\rho}$ are of order $O\left(  \rho^{2}\right)  $, and therefore
$G^{\rho\rho}\partial_{\rho}\left(  \left\vert W\right\vert ^{2}\right)  $ is
$O\left(  \rho^{3}\right)  $, whereas the determinant of the induced metric
$\sqrt{h}$ is of order $O\left(  \rho^{-5/2}\right)  $. Therefore, the
boundary term is of order $O\left(  \rho^{1/2}\right)  $ and it vanishes asymptotically.

\bibliographystyle{JHEP}
\bibliography{renvolbib2}

\end{document}